\documentclass[fleqn,usenatbib]{mnras}

\usepackage{newtxtext,newtxmath}
\usepackage[T1]{fontenc}

\DeclareRobustCommand{\VAN}[3]{#2}
\let\VANthebibliography\thebibliography
\def\thebibliography{\DeclareRobustCommand{\VAN}[3]{##3}\VANthebibliography}

\usepackage{graphicx}    % Including figure files
\usepackage{amsmath}    % Advanced maths commands
\usepackage{amssymb}    % Extra maths symbols

\title[Radio observations of ASASSN-16fp]{Radio view of a broad-line Type Ic supernova ASASSN-16fp}

\author[Nayana et al.]{
Nayana A.J.,$^{1,2}$\thanks{E-mail: nayana$_{-}$a@uaeu.ac.ae}
and Poonam Chandra$^{2}$
\\
$^{1}$Department of Physics, United Arab Emirates University, Al-Ain, UAE, 15551\\
$^{2}$National Centre for Radio Astrophysics, Tata Institute of Fundamental Research, PO Box 3, Pune, 411007, India\\
}

\date{Accepted XXX. Received YYY; in original form ZZZ}

\pubyear{2020}

\begin{document}
\label{firstpage}
\pagerange{\pageref{firstpage}--\pageref{lastpage}}
\maketitle

\begin{abstract}
We present extensive radio observations of a Type Ic supernova, ASASSN-16fp. Our data represents the lowest frequency observations of the SN beyond 1000 days with a frequency range of $0.33-25$ GHz and a temporal range of $\sim$ 8 to 1136 days post-explosion. The observations are best represented by a model of synchrotron emission from a shocked circumstellar shell initially suppressed by synchrotron self-absorption. Assuming equipartition of energy between relativistic particles and magnetic fields, we estimate the velocity and radius of the blast wave to be $v \sim 0.15c$ and $r \sim 3.4$ $\times$ $10^{15}$ cm respectively at $t_{0}$ $\sim$ 8 days post-explosion. We infer the total internal energy of the radio-emitting material evolves as $E$ $\sim$ 0.37 $\times$ $10^{47}$ (t/t$_{0}$)$^{0.65}$ erg. We determine the mass-loss rate of the progenitor star to be $\dot{M}$ $\sim$ $(0.4-3.2) \times10^{-5}$ $M_{\odot}\rm yr^{-1}$ at various epochs post-explosion, consistent with the mass-loss rate of Galactic Wolf-Rayet stars. The radio light curves and spectra show a signature of density enhancement in the CSM at a radius of $\sim$ $1.10 \times 10^{16}$ cm from the explosion center.
\end{abstract}


%%%%%%%%%%%%%%%%%%%%%%%%%%%%%%%%%%%%%%%%%%%%%%%%%%

%%%%%%%%%%%%%%%%% BODY OF PAPER %%%%%%%%%%%%%%%%%%

\section{Introduction}
 \label{sec:intro}
Core-collapse Supernovae (SNe) show considerable diversity in their observational signatures. They are classified into various sub-classes based on the distinct features in the optical light curve and spectra. Type Ib/c SNe (hereafter SNe Ib/c) is a sub-class of core-collapse SNe that shows no hydrogen lines in their optical spectra \citep{filippenko1997}. In a volume-limited sample of all core-collapse SNe, Type Ibc comprises of $\sim$22\% of the sample \citep{smith2011} and hence is an important mode of massive stellar death. Among SNe Ib/c , SNe Ib is characterized by the presence of helium lines in their spectra whereas SNe Ic does not show any helium lines (or very weak helium lines). Some SNe Ic show broad absorption lines in their optical spectra and they are called broad-lined SNe Ic \citep[SNe Ic-BL;][]{valenti2008}. SNe Ic-BL are understood to have higher energy than typical SNe Ib/c \citep{foley2003,valenti2008}. A sub-population of SNe Ic-BL are associated with Gamma-ray bursts (GRBs), some are SN\,1998bw/GRB\,980425 \citep{galama1998,pian2000}, SN\,2003dh/GRB\,030329 \citep{berger2003, hjorth2003, mazzali2003, frail2005}, SN\,2016jca/GRB\,161219B \citep{dai2016,deugarte2016,ashall2017,alexander2016,nayana2016b}. %There are few SNe Ic-BL (eg: SNe\,2009bb and 2012ap) that shows He-features in their spectra and they are known to be transitional SN between SNe Ib and SNe Ic-BL \citep{pignata2011,milisavljevic2015}. 

The progenitors of SNe Ib/c are understood to be massive stars that have lost their hydrogen and/or helium layers before the core-collapse. The two popular progenitor models are the following. A single massive star that lost its outer hydrogen and/or helium envelope via strong stellar winds \citep{ensman1988} or a massive star in a binary system where the outer stellar layers are stripped off due to binary interactions \citep{woosley1995,yoon2010,yoon2015}. There is one direct detection of the progenitor of this class in the case of SN\,PTF13bv \citep[a Type Ib SN;][]{cao2013}. The zero-age-main sequence (ZAMS) mass of the progenitor was determined to be $\sim$ 30 $M_{\odot}$ from broad-band magnitudes \citep{cao2013,groh2013}. Alternatively the progenitor colors can also be reproduced in a binary model with ZAMS mass combinations of either 20 + 19 $M_{\odot}$ or 10 + 8 $M_{\odot}$ \citep{bersten2014,eldridge2015}. The direct detections still lack the sensitivity to discriminate the single star and binary progenitor models. Other than this one detection, there is no direct evidence of progenitor stars of this class. Thus the progenitor scenario and the mass range of progenitor stars of SNe Ibc is still an open problem. Any information about the nature of the progenitor system via indirect probes is important.

Radio emission from the hydrodynamical interaction of supernova (SN) with the circumstellar medium (CSM) is an important probe to study the progenitor properties and immediate CSM \citep{chevalier1982b}. Radio observations and modeling can constrain various physical parameters like the mass-loss rate of the progenitor star, the radius of the blast wave, the post-shock magnetic field, and CSM density. The density of the CSM will be different for single mass progenitors and binary progenitors. A single Wolf-Rayet (WR) star will have a wind stratified media around it \citep{chu2002} whereas a binary system is likely to have a disrupted CSM due to the outflows in a common envelope phase \citep{podsiadlowski1992}. Radio light curves probe the density structure of the CSM and can discriminate between the two progenitor scenarios. 

In this paper, we present extensive radio observations of an SN Ic-BL, ASASSN-16fp over a frequency range of $0.33-25$ GHz and a temporal range of $\sim$ 8 to 1136 years post-explosion. We model the radio observation as synchrotron emission from the SN interaction with the CSM created by a steady stellar wind from the progenitor star \citep{chevalier1982a,chevalier1982b}. We derive the mass-loss rate of the progenitor star and blast wave parameters at multiple epochs of the SN evolution. We also compare the properties of ASASSN-16fp with other SNe Ib/c.

The organisation of the paper is the following. In \S \ref{sec:asassn16fp}, we review the previous studies done on ASASSN-16fp from the literature. We present the observations and data analysis in \S \ref{sec:obs}. The radio model is discussed in \S \ref{sec:modelling}. Our results are presented and discussed in \S \ref{sec:results and discussion} and we summarize the paper in \S \ref{sec:summary}.

\section{ASASSN-16fp}
\label{sec:asassn16fp}
ASASSN-16fp was discovered by the All-sky automated survey for the supernovae (ASAS-SN) team \citep{holoien2016} on 2016 May 27.6 (UT) in the nearby galaxy UGC\,11868 with an apparent magnitude of $\sim$ 15.7 (V-band). The distance towards the SN is $D = 17.2$ Mpc (from NED). The SN was initially classified as a SN\,Ic-BL \citep{elias2016} and later as a transitional SN between SNe Ib and SNe Ic-BL due to the presence of He lines in the early optical spectrum \citep{yamanaka2017}. The date of explosion of ASASSN-16fp was estimated to be 2016 May 24.5 (UT) by extrapolating the rising part of the optical light curve \citep{yamanaka2017}. \cite{brajesh2017} carried out optical follow-up observations of ASASSN-16fp during the photospheric phase (-10 to +33 days with respect to the B band maximum) and presented the light curve and low resolution spectra. 
\cite{brajesh2017} estimated the date of explosion to be 2016 May 25.9 (UT) by fitting the good cadence data points from the pre-maximum phase. The last non-detection of ASASSN-16fp was on 2016 May 21.5 (UT) \citep{holoien2016}. Thus both the explosion dates derived by \cite{yamanaka2017} and \citep{brajesh2017} are consistent with a maximum uncertainty of $\Delta$ t = May $27.6-21.5$ = 6.1. In this work, we adopt the date of explosion to be 2016 May 25.9 (UT).

\cite{prentice2018} presented optical observations of ASASSN-16fp from 2 to 450 days post-explosion and analyzed the physical properties. The early photospheric phase spectra showed the presence of helium in a C/O dominated shell. The authors derived the mass of the ejected material from the SN to be $\sim$ $2.5-4$ $M_{\odot}$ with a kinetic energy of $\sim$ ($4.5-7$) $\times$ $10^{51}$ ergs. They estimated a progenitor mass of $23-28$ $M_{\odot}$ with almost completely stripped hydrogen and helium layers.

X-ray emission was detected from ASASSN-16fp with the X-ray telescope (XRT) onboard \textit{Swift} satellite \citep{burrows2005} on 2016 May 27.7 with a flux  of 8.7$^{+4.6}_{-3.5}$ $\times$ 10$^{-14}$ erg\,s$^{-1}$\,cm$^{-2}$ in the of $0.3-10.0$ keV band \citep{grupe2016}. Radio emission was detected at 15 GHz with the Arcminute microkelvin imager (AMI) Large Array \citep{zwart2008} between $28-31$ May 2016 with a flux density of 1.4 mJy on 28 May 2016 \citep{mooley2016}. \cite{argo2016} detected radio emission at 5 GHz with the enhanced multi element remotely linked interferometer network (e-MERLIN) with a flux density of 1.3 $\pm$ 0.2 mJy on 2016 June 5.92, resulting a 5 GHz spectral luminosuity of 5 $\times$ 10$^{26}$ erg\,s$^{-1}$\,Hz$^{-1}$. At low frequency, radio emission  was detected at 1.4 GHz with the Giant Metrewave Radio Telescope \citep[GMRT;][]{swarup1991} with a flux density of 0.25 mJy on 2016 June 28.8 \citep{nayana2016}.

\cite{terreran2019} reported the multi-wavelength observations of ASASSN-16fp from $\gamma$-rays to radio wavelengths. The authors derived the ejecta mass and kinetic energy of the SN to be $M_{\rm ej}$ $\sim$ $4-7$ $M_{\odot}$ and $E_{\rm k}$ $\sim$ $7-8$ $\times$ $10^{51}$ erg respectively from bolometric light curve modeling. The mass-loss rate of the progenitor star was estimated as $\dot{M}$ = ($1-2$) $\times$ 10$^{-4}$ $M_{\odot}$\,yr$^{-1}$ from X-ray observations \citep{terreran2019}. 

\section{Observations and Data Analysis}
\label{sec:obs}
\subsection{GMRT Observations}
\label{sec:gmrt-obs}
\begin{figure*}
\begin{centering}
\includegraphics[scale=0.75]{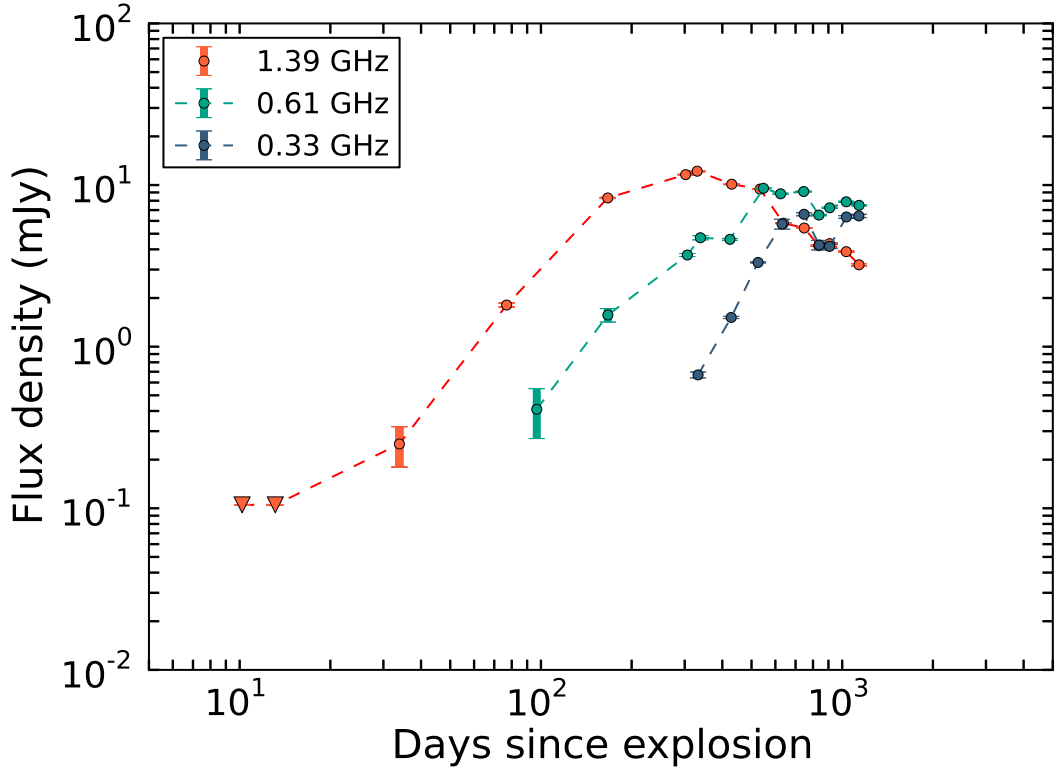}
 \caption{GMRT light curves of ASASSN-16fp at frequencies 0.325, 0.610 and 1.40 GHz. The days since explosion are calculated assuming the date of explosion as 2016 May 25.9 (UT). The inverted triangle denotes 3$\sigma$ upper limits. Majority of the error bars in the figure (see table \ref{tab:gmrt}) are smaller than the marker size.}
    \label{fig:lc-gmrt}
    \end{centering}
\end{figure*}

We started observing ASASSN-16fp with the GMRT since 2016 Jun 05.09 (UT) ($\sim$ 10 days post-explosion) till 2019 Jul 06.87 ($\sim$ 1136 days post-explosion) at 1390, 610 and 325 MHz. Data were collected in the full intensity mode with an integration time of 16.1 sec. We used an observing bandwidth of 33 MHz split into 256 channels at all three frequencies. 3C286, 3C48 and 3C147 were used as the flux calibrators. We used J2139+143 and J2251+188 as phase calibrators. The data were analyzed using the Astronomical Image Processing System \citep[AIPS;][]{greisen2003} using standard techniques. Initial flagging and calibration were done using the software FLAGCAL, developed for automatic flagging and calibration for the GMRT data \citep{prasad2012}. The calibrated data were imaged using AIPS task IMAGR. The flux density and errors are obtained from gaussian fit to the SN position using the task JMFIT. The details of GMRT observations and the flux densities of the SN are presented in Table \ref{tab:gmrt}. The GMRT radio light curves at 1.39, 0.61 and 0.33 GHz spanning $\sim$ $10-1136$ days post-explosion are shown in Fig. \ref{fig:lc-gmrt}. 
\subsection{JVLA observations}
We observed ASASSN-16fp with the Karl G. Jansky Very Large Array (JVLA) on 2017 Feb 17.65, 17.67 and Feb 25.8 (project code 17A-167) spanning a frequency range 2.2 to 9.7 GHz. The observations were done in the standard continuum mode with a bandwidth of 2 GHz split into 16 spectral windows. We used 3C286 and 3C48 as the flux density calibrators and J2139+143 as the phase calibrator.

\begin{figure*}
\begin{centering}
\includegraphics[scale=0.75]{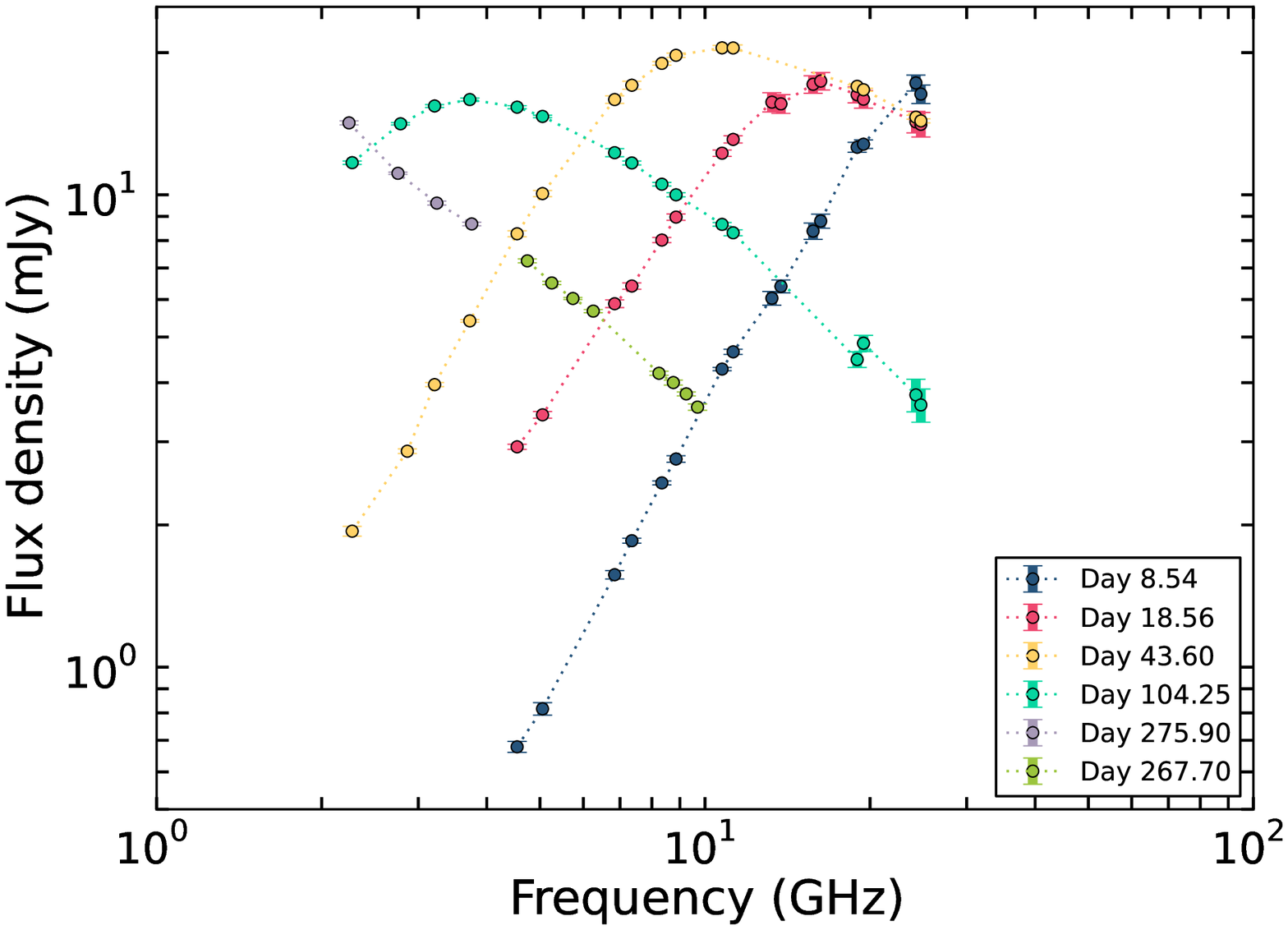}
 \caption{JVLA spectra of ASASSN-16fp on day 8 to 276 days post explosion spanning a frequency range frequencies $\sim$ $2-24$ GHz. The days since explosion are calculated assuming the date of explosion as 2016 May 25.9 (UT). The error bars in the figure (see table \ref{tab:vla}) are smaller than the marker size.}
    \label{fig:spectra-vla}
    \end{centering}
\end{figure*}

We also analyzed the publicly available archival JVLA data of ASASSN-16fp at five epochs from 2016 June 03.44 (UT) to 2016 Sep 07.15 (UT), spanning a frequency range $2-25$ GHz. The JVLA observations at each frequency were carried out with a bandwidth of $\approx$ 2 GHz split into 16 spectral windows. 3C286 and 3C48 were observed for the flux density calibration, and J2139+143 was observed as the phase calibrator. The data analysis was done using standard packages within the Common Astronomy Software Applications package \citep[CASA;][]{mcmullin2007}. We split the data into four sub-bands, each of $\sim$ 0.5 GHz bandwidth during the data reduction. The details of VLA observations and the flux densities of the SN at various epochs are summarised in Table \ref{tab:vla}. We plot the full VLA dataset in Fig. \ref{fig:spectra-vla}. 

\begin{table*}
    \centering
    \caption{Details of GMRT observations of ASASSN-16fp.}
    \label{tab:gmrt}
    \begin{tabular}{lccccc} % four columns, alignment for each
        \hline
        Date of Observation & Age$^{a}$ & Frequency & Flux density & rms\\
        (UT) & (Day) & (GHz) & (mJy)$^{b}$  & ($\mu$Jy\,beam$^{-1}$) \\
        \hline
    2016  Jun  05.09    &    10.19        &    1.39    &     $<$0.12    &  40     \\ 
2016  Jun  08.03    &    13.13        &    1.39    &    $<$ 0.12    &  40     \\ 
2016  Jun  28.80    &    33.90        &    1.39    &     0.25     $\pm$    0.09 & 70   \\ 
2016  Aug  10.85    &    76.95        &    1.39    &     1.81     $\pm$    0.08  & 50   \\ 
2016  Nov  08.55    &    166.65      &      1.39    &     8.32     $\pm$    0.14  & 70    \\
2017  Mar  24.05    &    302.15      &      1.39    &     11.60     $\pm$    0.13  & 50   \\ 
2017  Apr  21.01    &    330.11      &      1.39    &     12.18     $\pm$    0.16  & 90    \\
2017  Jul  29.72    &    429.82      &      1.39    &     10.12     $\pm$    0.11  & 60     \\
2017  Nov  10.49    &   533.59      &    1.39    &     9.44   $\pm$    0.06 & 40     \\   
2018  Feb  23.09    &   638.19      &    1.39    &     5.81   $\pm$    0.05 & 35     \\ 
2018  Jun  12.84    &   747.94      &    1.39    &     5.42   $\pm$    0.06 & 38     \\ 
2018  Sep  08.71    &   835.81      &    1.39    &     4.21  $\pm$    0.06  &  35     \\ 
2018  Nov  20.38    &   908.48      &     1.39    &     4.34    $\pm$  0.05  &  45   \\ 
2019 Mar 23.24      &   1031.34     &  1.39   &  3.87  $\pm$ 0.05  & 38     \\
2019 Jul 06.76      &   1136.86     &  1.39   &  3.21 $\pm$ 0.07   & 53   \\
\hline 
2016  Aug  30.76    &    96.86        &    0.61    &     0.41     $\pm$    0.16  &  140     \\ 
2016  Nov  08.71    &    166.81      &      0.61    &     1.57     $\pm$    0.17  &  150     \\ 
2017  Mar  28.04    &    306.14      &      0.61    &     3.69     $\pm$    0.11  &  80     \\ 
2017  Apr  29.18    &    338.28      &      0.61    &     04.72     $\pm$    0.16  &  140     \\ 
2017  Jul  23.91    &     424.01    &    0.61    &     4.61   $\pm$   0.09 &  60     \\ 
2017  Nov  24.63    &     547.73    &    0.61    &     9.55   $\pm$   0.12 &  90     \\ 
2018  Feb  09.37    &     624.47    &    0.61    &     8.83   $\pm$   0.14 &  85     \\ 
2018  Jun  09.91    &     745.01    &    0.61    &     9.12   $\pm$   0.11 &  65     \\ 
2018  Sep  09.71    &     836.81    &    0.61    &     6.52   $\pm$   0.11 &  65     \\ 
2018  Nov  20.54    &     908.64    &    0.61    &     7.23    $\pm$   0.13 & 65     \\ 
2019 Mar 23.05      &     1031.15   &   0.61    &    7.88   $\pm$    0.09  & 68     \\
2019 Jul 06.87      &     1136.97   &   0.61   &    7.49  $\pm$ 0.18      & 67     \\
\hline
2017  Apr  23.14    &    332.24      &       0.325    &     0.67    $\pm$    0.31  &  280     \\ 
2017  Jul  27.92    &     428.02    &    0.325    &     1.52   $\pm$    0.28 &  240     \\ 
2017  Nov  02.55    &     525.65    &    0.325    &     3.32   $\pm$    0.29 & 240     \\ 
2018  Feb  16.39    &     631.49    &    0.325    &     5.74   $\pm$    0.42 & 400     \\ 
2018  Jun  11.91    &     747.01    &    0.325    &     6.59   $\pm$    0.21 &  142     \\ 
2018  Sep  11.56    &     838.66    &    0.325    &     4.26   $\pm$    0.34   & 290     \\ 
2018  Nov  19.57    &     907.67    &     0.325    &     4.17   $\pm$    0.25  & 130    \\
2019  Mar  24.31    &     1032.41   &    0.325  &   6.35    $\pm$   0.13   & 118  \\
2019 Jul 05.76     &     1135.86    &   0.325   &   6.44   $\pm$  0.30   & 155 \\
        \hline
\end{tabular}\\
\parbox{110mm}{
$a$ The age is calculated assuming 2016 May 25.9 (UT) as the date of explosion (see \S \ref{sec:intro}).\\
$b$ The errors in the flux density are from the task JMFIT.
}
\end{table*}

\begin{table*}
    \centering
    \caption{Details of JVLA observations of ASASSN-16fp.}
    \label{tab:vla}
    \begin{tabular}{lcccccc} % four columns, alignment for each
        \hline
Date of Observation & Age$^{a}$ & Frequency & VLA Array & Flux density & rms\\
(UT) & (Day) & (GHz) & configuration & (mJy)$^{b}$  & ($\mu$Jy\,beam$^{-1}$) \\
        \hline
    2016 Jun 03.44  & 8.54  & 4.543  & B  & 0.678$\pm$0.038  & 23   \\
         - & -  & 5.055 & B  & 0.816$\pm$0.035 & 22  \\    
         - & -  & 6.843 & B  & 1.569$\pm$0.032  & 18  \\
         - & -  & 7.355 & B  & 1.852$\pm$0.022  & 19  \\
         - & -  & 8.343 & B  & 2.455$\pm$0.024  & 20  \\
         - & -  & 8.855 & B  & 2.758$\pm$0.044  & 20  \\
          - & -  & 10.743 & B  & 4.276$\pm$0.033  & 28  \\
          - & -  & 11.255 & B  & 4.651$\pm$0.057  & 28  \\
          - & -  & 13.243 &  B & 6.04$\pm$0.200  & 44  \\
           - & -  & 13.755 &  B & 6.400$\pm$0.200  & 53  \\
           - & -  & 15.743 &  B & 8.380$\pm$0.330  & 79  \\
            - & -  & 16.255 & B  & 8.800$\pm$0.300  & 92  \\
            - & -  & 18.943 & B  & 12.609$\pm$0.306  & 139  \\
           - & -  & 19.455 &  B & 12.798$\pm$0.267  & 126  \\ 
            - & -  & 24.243 & B  & 17.251$\pm$0.668  & 334  \\
             - & -  & 24.755 & B  & 16.341$\pm$0.742  & 371  \\ 
2016 Jun 13.46  & 18.56  & 4.543  & B  & 2.927$\pm$0.037  & 22   \\
              - & -  & 5.055 & B  & 3.421$\pm$0.055  & 23  \\
               - & -  & 6.843 & B  & 5.880$\pm$0.110  & 21  \\
                - & -  & 7.355 & B  & 6.410$\pm$0.100  & 24  \\
                 - & -  & 8.343 & B  & 8.020$\pm$0.100  & 39  \\
                  - & -  & 8.855 & B  & 8.970$\pm$0.140  & 42  \\
                   - & -  & 10.743 & B  & 12.250$\pm$0.190  & 72  \\
                    - & -  & 11.255 & B  & 13.100$\pm$0.230  & 66  \\
                     - & -  & 13.243 & B  & 15.710$\pm$0.730  & 147  \\
                      - & -  & 13.755 & B  & 15.580$\pm$0.710  & 140  \\
               - & -  & 15.743 & B  & 17.140$\pm$0.730  & 192  \\        
                - & -  & 16.255 & B  & 17.410$\pm$0.740  & 228  \\    
                 - & -  & 18.943 & B  & 16.290$\pm$0.630  & 242  \\
                  - & -  & 19.455 & B  & 15.940$\pm$0.690  & 236  \\
                   - & -  & 24.243 & B  & 14.270$\pm$0.750  & 294  \\     
                    - & -  & 24.755 & B  & 14.100$\pm$0.840  & 306  \\  
2016 July 08.50  & 43.6  & 2.273  & B  & 1.940$\pm$0.046  & 40   \\    
                - & -  & 2.865 & B  & 2.865$\pm$0.032  & 30  \\  
                 - & -  & 3.213 & B  & 3.964$\pm$0.039  & 26  \\ 
                 -  & -  & 3.725 & B  & 5.408$\pm$0.034  & 23  \\   
                 - & -  & 4.543 &  B & 8.270$\pm$0.120  & 27  \\   
                  - & -  & 5.055 & B  & 10.060$\pm$0.150  & 29  \\  
                - & -  & 6.843 & B  & 15.910$\pm$0.290  & 45  \\    
               - & -  & 7.355 &  B & 17.060$\pm$0.330  & 49  \\ 
                - & -  & 8.343 & B  & 18.980$\pm$0.170  & 74  \\  
                 - & -  & 8.855 & B  & 19.740$\pm$0.210  & 97  \\  
                  - & -  & 10.743 & B  & 20.470$\pm$0.250  & 93  \\ 
                   - & -  & 11.255 & B  & 20.460$\pm$0.300  & 110  \\
                    - & -  & 18.943 & B  & 16.970$\pm$0.140  & 111  \\
                 - & -  & 19.455 & B  & 16.690$\pm$0.120  & 113  \\  
                  - & -  & 24.243 & B  & 14.600$\pm$0.138  & 133  \\  
                   - & -  & 24.755 & B  & 14.340$\pm$0.140  & 120  \\
2016 Sep 07.15  & 104.25  & 2.273 &  A & 11.697$\pm$0.097  & 65  \\    
                 - & -  & 2.785 & A  & 14.138$\pm$0.080  & 35  \\ 
                  - & -  & 3.213 &  A & 15.420$\pm$0.110  & 29  \\ 
                 - & -  & 3.725 & A  & 15.910$\pm$0.120  & 28  \\  
                  - & -  & 4.543 & A  & 15.330$\pm$0.110  & 24  \\ 
                 - & -  & 5.055 & A  & 14.654$\pm$0.088  & 26  \\
                  - & -  & 6.843 &  A & 12.280$\pm$0.240  & 30  \\  
                  - & -  & 7.355 &  A & 11.680$\pm$0.110  & 27  \\ 
                 - & -  & 8.343 & A  & 10.530$\pm$0.094  & 31  \\  
                - & -  & 8.855 & A  & 10.000$\pm$0.100  & 35  \\  
        \hline
\end{tabular}\\
\parbox{110mm}{
$a$ The age is calculated assuming 2016 May 25.9 (UT) as the date of explosion (see \S \ref{sec:intro}).\\
$b$ The errors in the flux density are from the task JMFIT.
}
\end{table*}

\begin{table*}
    \centering
    \contcaption{Details of JVLA observations of ASASSN-16fp.}
    \label{tab:vla-continued}
    \begin{tabular}{lcccccc} % four columns, alignment for each
        \hline
        Date of Observation & Age$^{a}$ & Frequency & VLA Array & Flux density & rms\\
        (UT) & (Day) & (GHz) & configuration & (mJy)$^{b}$  & ($\mu$Jy\,beam$^{-1}$) \\
        \hline
2016 Sep 07.15  & 104.25  & 10.743 & A  & 8.657$\pm$0.076  & 37  \\
                 - & -  & 11.255 & A  & 8.310$\pm$0.120  & 35  \\  
                  - & -  & 18.943 & A  & 4.480$\pm$0.170  & 94  \\    
                 - & -  & 19.455 &  A & 4.850$\pm$0.190  & 97  \\
                  - & -  & 24.243 & A  & 3.770$\pm$0.300  & 145  \\
                   - & -  & 24.755 &  A & 3.590$\pm$0.290  & 144  \\ 
2017 Feb 25.80  & 275.90  & 2.243 & D  & 14.200$\pm$0.140 & 139  \\   
                 - & -  & 2.755 & D  & 11.105$\pm$0.059  & 56  \\ 
                 - & -  & 3.243 & D  & 9.604$\pm$0.091   & 46 \\
                 - & -  & 3.755 & D  & 8.676$\pm$0.076   & 45 \\ 
2017 Feb 17.67  & 267.77  & 4.743 & D  & 7.247$\pm$0.072  & 45  \\
                 - & - & 5.255  & D  & 6.507$\pm$0.051 & 31 \\
                 - & - & 5.743  & D  & 6.032$\pm$0.030 & 29 \\
                 - & - & 6.255  & D  & 5.672$\pm$0.043 & 28 \\
2017 Feb 17.65  & 267.75  & 8.243 & D  & 4.189$\pm$0.043  & 22  \\
                 - & - & 8.755 & D & 4.004$\pm$0.049  & 20 \\
                 - & - & 9.243 & D & 3.789$\pm$0.036  & 25 \\
                 - & - & 9.692 & D & 3.552$\pm$0.056  & 24 \\ 
        \hline
\end{tabular}\\
\parbox{110mm}{
$a$ The age is calculated assuming 2016 May 25.9 (UT) as the date of explosion (see \S \ref{sec:intro}).\\
$b$ The errors in the flux density are from the task JMFIT.
}
\end{table*}

\section{A Radio Model}
\label{sec:modelling}
Radio emission from core-collapse SNe is synchrotron in origin \citep{chevalier1982a,chevalier1982b}, produced due to the interaction of SN ejecta with the CSM created by the stellar wind of the progenitor star. According to the standard model, the hydrodynamical evolution of the interaction region is self-similar across the shock discontinuity producing a shock wave of radius $R(t) \propto t^{m}$ \citep{chevalier1982a}. For an outer ejecta density profile of $\rho_{\rm ej} \propto r^{-n}$ and CSM density of $\rho_{\rm csm}$ $\propto$ $r^{-s}$, the shock deceleration parameter $m=(n-3)/(n-s)$. The model also assumes that a fixed fraction of shock energy is fed into the relativistic particle energy density and magnetic field energy density \citep[model 1 of;][]{chevalier1996}. At an early time, there are different absorption processes that supress radio synchrotron emission. 
It could be either free-free absorption (FFA) by the ionized CSM or synchrotron self-absorption (SSA) by the same electron population that produces radio emission \citep{chevalier1982b,chevalier1998}. The early radio light curve shows the evolution of the SN in the optically thick regime where the absorption processes are dominant. Later, the optical depth decreases as the shell expands and when the optical depth becomes unity, the light curve shows the transition from optically thick to thin regime. 

We model the radio data of ASASSN-16fp with the standard model \citep{chevalier1982b}. We fit the data with both FFA \citep{chevalier1982b,weiler2002} and SSA models \citep{chevalier1998} following the procedure similar to \cite{nayana2018}.
For FFA model, the radio flux density, $F(\nu,t)$ is 

\begin{equation}
F(\nu,t)=K_{1} \left( \frac{\nu}{5\hspace{0.1 cm} \rm GHz}\right)^{\alpha} \left( \frac{t}{10\, \rm day}\right)^{\beta}e^{-\tau_{\rm ffa}(\nu,t)}
\end{equation}

Where $\tau_{\rm ffa}(\nu,t)$ is the free-free optical depth due to the ionized CSM defined as.

\begin{equation}
\tau_{\rm ffa}(\nu,t)=K_{2}\left(\frac{\nu}{5\hspace{0.1 cm}\rm GHz}\right)^{-2.1}\left(\frac{t}{10\, \rm day}\right)^{\delta}
\end{equation}

where $K_{1}$ and $K_{2}$ are the flux density and optical depth normalization parameters. $\alpha$ and $\beta$ denotes the spectral and temporal indices of the radio flux densities. 
%The spectral index $\alpha$ can be related to the electron energy index $p$ as $\alpha=-(p-1)/2$ where $p$ is defined such that $N(E)\propto E^{-p}$. $\delta$ denotes the temporal index of optical depth and is related to shock deceleration parameter $m$ ($R \propto$ $t^{m}$) as $m=-\delta/3$ for $s=2$. 
For SSA model, the radio flux density is \citep{chevalier1998}
\begin{equation}
F(\nu,t)=K_{1} \left( \frac{\nu}{5\, \rm GHz}\right)^{2.5} \left( \frac{t}{10\, \rm day}\right)^{a} \left(1- e^{-\tau_{\rm ssa}(\nu,t)} \right) 
\end{equation}

Where $\tau_{\rm ssa}(\nu,t)$ is the SSA optical depth given by

\begin{equation}
\tau_{\rm ssa}(\nu,t)=K_{2}\left(\frac{\nu}{\rm 5\, GHz}\right)^{-(p+4)/2}\left(\frac{t}{10\,\rm day}\right)^{-(a+b)}
\end{equation}

Where $a$ and $b$ denotes the temporal index of flux densities in the optically thick ($F_{\nu} \propto t^{a}$) and thin phase ($F_{\nu} \propto t^{-b}$). For model 1 of \cite{chevalier1996}, $a$, $b$ and $p$ can be related to the shock deceleration parameter $m$ as the following. $a=2m + 0.5$ in the optically thick phase and $b=(p+5-6m)/2$ in the optically thin phase. 

We carry out a two-variable fit $F(\nu,t)$ to the complete radio data with FFA and SSA models. The free parameters in the FFA model are $K_{1}$, $K_{2}$, $\alpha$, $\beta$ and $\delta$ and in the SSA model are $K_{1}$, $K_{2}$, $a$, $b$ and $p$. We have a total of 112 flux density measurements at multiple epochs and frequencies. With 5 free parameters, the fit has 107 degrees of freedom. We use the chi-square minimization algorithm available in python-scipy \citep{virtanen2019}. In the fitting routine, 10\% of flux density is added in quadrature to the JMFIT errors as systematic uncertainty to account for the calibration errors. The maximum calibration errors in various frequency bands of the VLA is $(3-10)$\% \citep{weiler1986}. The calibration error at multiple bands of GMRT is $\sim$10\% \citep{chandra2017}. 

\section{Results}
\label{sec:results and discussion}
\begin{figure*}
\begin{centering}
    \includegraphics*[width=0.9\textwidth]{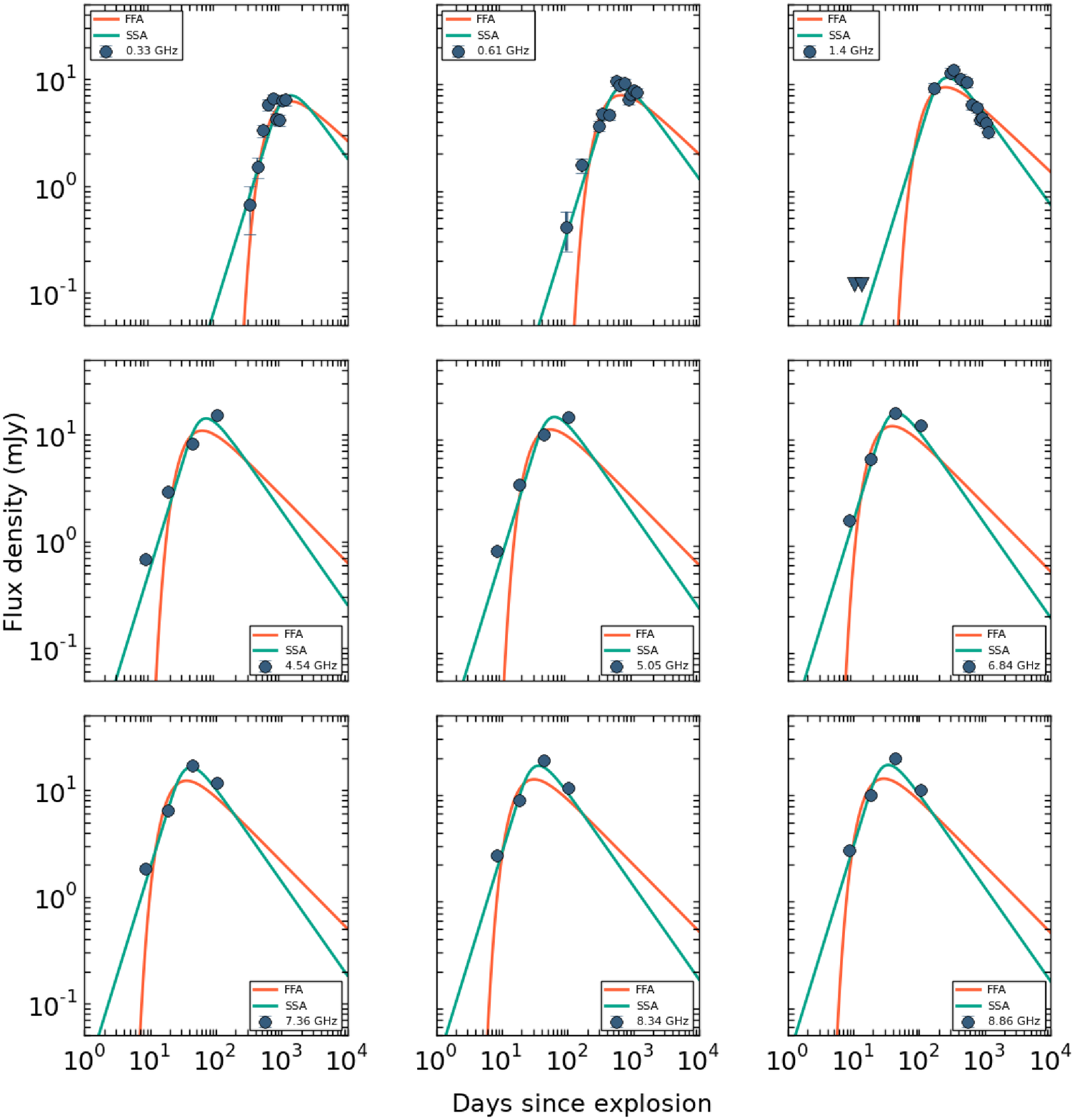}
\caption{SSA and FFA model fits to the radio light curves of ASASSN-16fp at 0.33, 0.61, 1.40, 4.54, 5.05, 6.84, 7.36, 8.34 and 8.86 GHz. Green solid line denotes the SSA model and red solid line denotes the FFA model. The days since explosion is calculated assuming the date of explosion as 2016 May 25.9 (UT). Majority of the error bars in the figure (see table\ref{tab:gmrt} and \ref{tab:vla}) are smaller than the marker size.} 
\label{fig:lc-fit-one}
\end{centering}
\end{figure*}

\begin{figure*}
\begin{centering}
    \includegraphics*[width=0.9\textwidth]{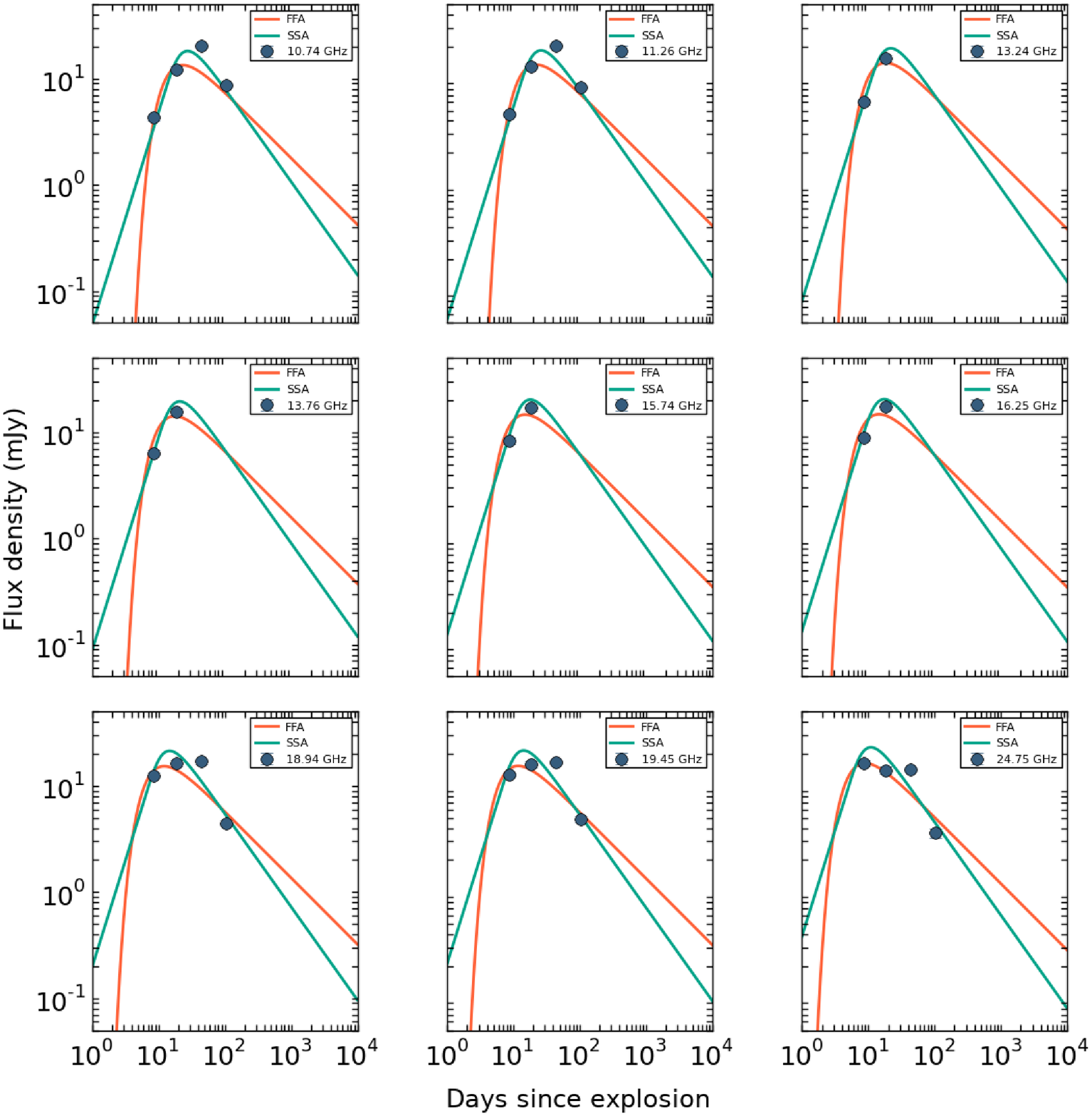}
\caption{SSA and FFA model fits to the radio light curves of ASASSN-16fp at 10.74, 11.26, 13.24, 13.76, 15.74, 16.25, 18.94, 19.45 and 24.75 GHz. Green solid line denotes the SSA model and red solid line denotes the FFA model. The days since explosion is calculated assuming the date of explosion as 2016 May 25.9 (UT). The error bars in the figure (see table \ref{tab:vla}) are smaller than the marker size.}
\label{fig:lc-fit-two}
\end{centering}
\end{figure*}

\begin{figure*}
\begin{centering}
    \includegraphics*[width=0.9\textwidth]{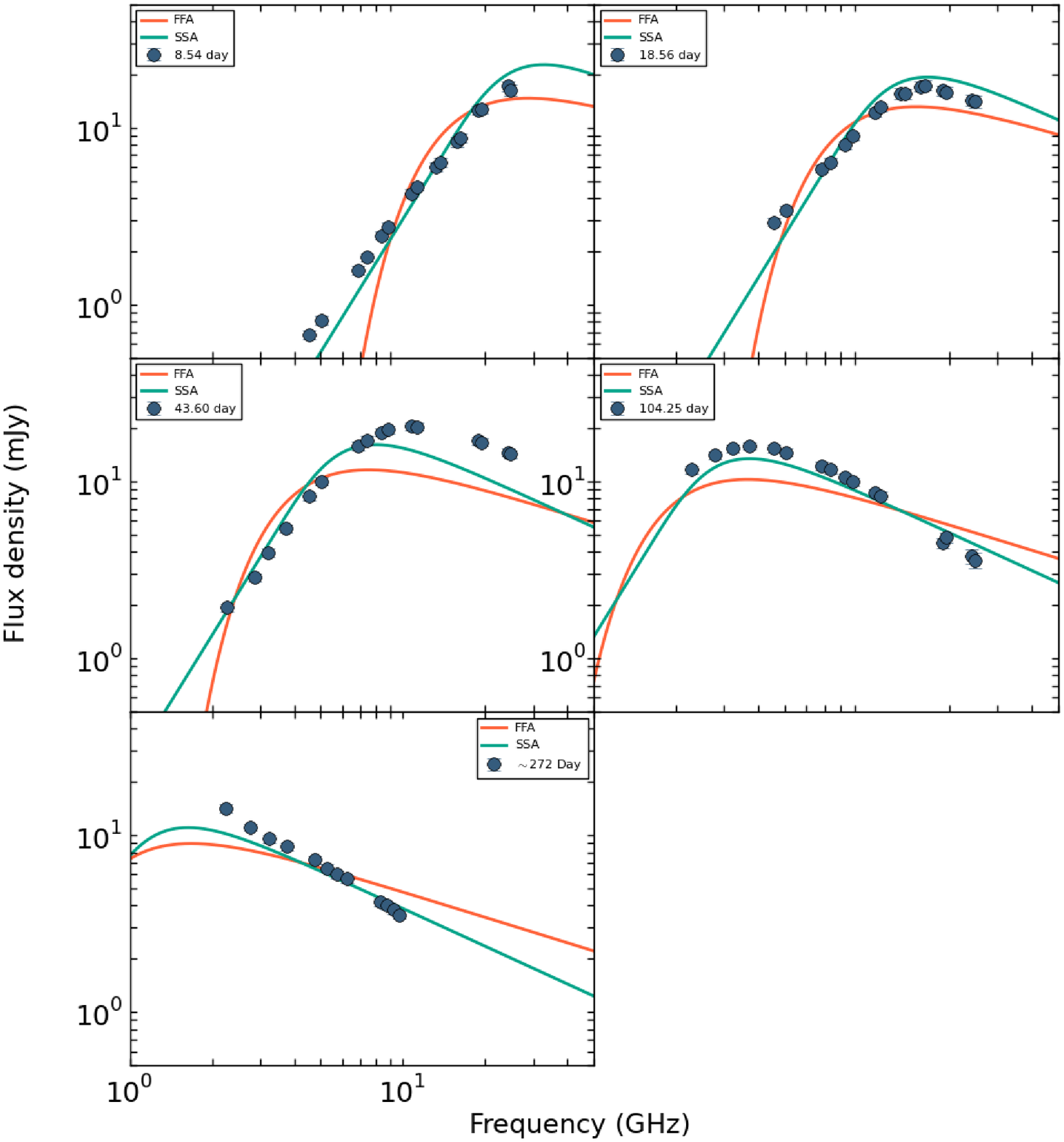}
\caption{SSA and FFA model fits to the radio spectra of ASASSN-16fp on day  8.54, 18.56, 43.60, 104.25, 267.70 and $\sim$ 272 post explosion. Green solid line denotes the SSA model and red solid line denotes the FFA model. The days since explosion is calculated assuming the date of explosion as 2016 May 25.9 (UT). The error bars in the figure (see table \ref{tab:vla}) are smaller than the marker size.}
\label{fig:spectra-fit}
\end{centering}
\end{figure*}

The best-fit parameters and the reduced-chi square values are presented in Table \ref{tab:fittedpara-global}. The best fit modeled light curves and spectra along with the observed data are shown in Fig. \ref{fig:lc-fit-one}, \ref{fig:lc-fit-two} and \ref{fig:spectra-fit}. The reduced chi-square value becomes higher for both FFA ($\chi_{\mu}^{2}$ = 62.0) and SSA ($\chi_{\mu}^{2}$ = 19.2) models if the fitting routine takes only 3\% of systematic error instead if 10\%. However, the best fit parameters are roughly same (within 20\%). From the reduced chi-square values and the fitted lightcurves and spectra, it is evident that the SSA model fits the data better than FFA. This is expected for SNe Ic since the plausible WR progenitors have fast stellar winds (few 1000 km\,s$^{-1}$) creating a less dense CSM. 

While overall data are better represented by a SSA model, it is still not a very good fit. There are a few deviations from the best fit model. In the light curve at 10.74, 11.26, 18.94, 19.45 and 24.75 GHz, the flux density measurement at 43.6 days post-explosion is slightly above the model prediction. The trend is more evident in the spectral fit (see Fig \ref{fig:spectra-fit}) where all flux density points above $\sim$ 10 GHz is slightly above the model prediction. We discuss this behavior in terms of a possible density enhancement in the CSM in \S \ref{sec:density-variations}.

\begin{table}
    \centering
    \caption{Best fit parameters for FFA and SSA model fits to ASASSN-16fp.}
    \label{tab:fittedpara-global}
    \begin{tabular}{cccccc} % four columns, alignment for each
        \hline
        FFA & SSA \\
        \hline
$K_{1}$ = 41.89 $\pm$ 6.38  &   $K_{1}$ = 0.72 $\pm$ 0.03   \\
$K_{2}$ = 7.41 $\pm$ 0.65 &  $K_{2}$ = 134.55 $\pm$ 18.88  \\
$\alpha$ = $-$0.47 $\pm$ 0.07  & $a$ = 1.99 $\pm$ 0.02  \\
$\beta$ = $-$0.58 $\pm$ 0.05  & $b$ = 0.85 $\pm$ 0.04   \\
$\delta$ = $-$1.80 $\pm$ 0.04 & $p$ = 2.40 $\pm$ 0.10 \\
$\chi_{\mu}^{2}$ = 9.83 & $\chi_{\mu}^{2}$ =  3.63 \\
d.o.f             = 107  &  d.o.f   = 107 \\
        \hline
\end{tabular}\\
\parbox{55mm}{
$K_{1}$ and $K_{2}$ are the normalization parameters of flux density and optical depth, respectively. In the FFA model, $\alpha$ and $\beta$ denotes the spectral and temporal evolution of the radio flux density. In the SSA model, $a$ and $b$ denotes the temporal index of flux density in the optically thick and thin regime respectively. $p$ denotes the electron energy index.
}
\end{table}

\subsection{Blast-wave parameters}
\label{sec:blastwave-para}
We model the single epoch spectra with the standard SSA model to derive the blast wave radius and post-shock magnetic field at multiple epochs \citep{chevalier1998}. 

For a power law electron distribution of the form $N(E)$ $\sim$ $E^{-p}$, the self absorbed synchrotron flux density is,

\begin{equation}
F_{\nu}\propto\frac{R^{2}}{D^{2}} B^{-1/2} \nu^{5/2} \hspace{1 cm} (\rm optically \hspace{0.25 cm} thick)
\label{eqn:ssa-optically-thick}
\end{equation} 

\begin{equation}
F_{\nu}\propto\frac{ f R^{3}}{D^{2}} N_{0} B^{(p+1)/2} \nu^{-(p-1)/2} \hspace{1 cm} (\rm optically \hspace{0.25 cm} thin)
\label{eqn:ssa-optically-thin}
\end{equation} 

Where $R$ is the radius of the blast wave, $D$ is the distance to the SN from the observer, $B$ is the magnetic field strength and $f$ is the volume filling factor of the radio emitting region. SSA defines a spectral break frequency ($\nu_{\rm p}$) below which the spectral evolution of the flux density is $\nu^{5/2}$ and above the spectral evolution is $\nu^{-(p-1)/2}$. At $\nu$ = $\nu_{\rm p}$ the two power laws (equation \ref{eqn:ssa-optically-thick} and \ref{eqn:ssa-optically-thin}) intersect and the corresponding flux density is $F_{\rm p}$. At this point equation, \ref{eqn:ssa-optically-thick} and \ref{eqn:ssa-optically-thin} can be inverted to obtain $R$ and $B$ assuming energy equipartition between the magnetic fields and relativistic particles. $R_{\rm p}$ and $B_{\rm p}$ for an electron power--law index $p$ is given by \cite{chevalier1998,chevalier2017} as.

\begin{eqnarray}
\label{eqn:Rp-general}
R_{\rm p} = \left[ \frac{6 c_{6}^{p+5} F_{\rm p}^{p+6}D^{2p+12}}{f_{\rm eB} f (p-2) \pi^{p+5} c_{5}^{p+6} E_{\rm l}^{p-2}} \right]^{\frac{1}{(2p+13)}} \left( \frac{\nu}{2c_{1}} \right)^{-1} 
\end{eqnarray}

\begin{eqnarray}
\label{eqn:Bp-general}
B_{\rm p} = \left[ \frac{36 \pi^{3} c_{5}}{f_{\rm eB}^{2}f^{2}(p-2)^{2}c_{6}^{3} E_{\rm l}^{2(p-2)} F_{\rm p} D^{2}}\right]^{\frac{2}{(2p+13)}} \left( \frac{\nu}{2c_{1}} \right)
\end{eqnarray}

 In the above equations, $f_{\rm eB}$ denotes the ratio of particle energy density to magnetic field energy density. Assuming energy equipartition between relativistic particles and magnetic fields, we take $f_{\rm eB}$=1. The constants $c_{5}$ and $c_{6}$ are tabulated as a function of $p$ in \cite{pacholczyk1970}. $c_{1} = 6.265 \times 10^{18}$ in CGS units \citep{chevalier2017}. $E_{\rm l}$ denotes the electron rest mass energy, i.e. 0.51 MeV.
 
\begin{figure}
\begin{centering}
    \includegraphics*[width=0.45\textwidth]{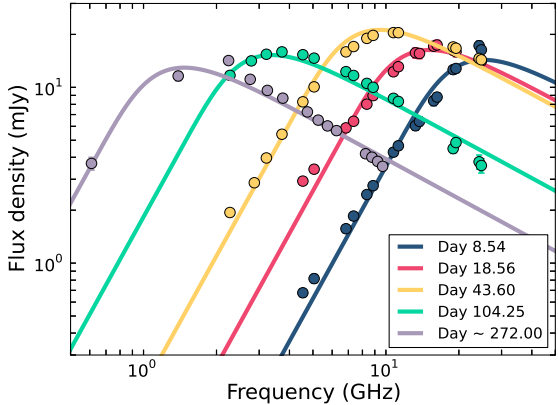}
\caption{Synchrotron self-absorption model fits to the single epoch radio data of ASASSN\,16fp. The epochs are day 8.54, 18.56, 43.60, 104.25, 272.00 post explosion. We assume the date of explosion to be 2016 May 25.9 (UT). The error bars in the figure (see table \ref{tab:vla}) are smaller than the marker size.}
\label{fig:spectra-single-epoch-fit}
\end{centering}
\end{figure}

\begin{figure*}
\begin{centering}
    \includegraphics*[width=5.8cm]{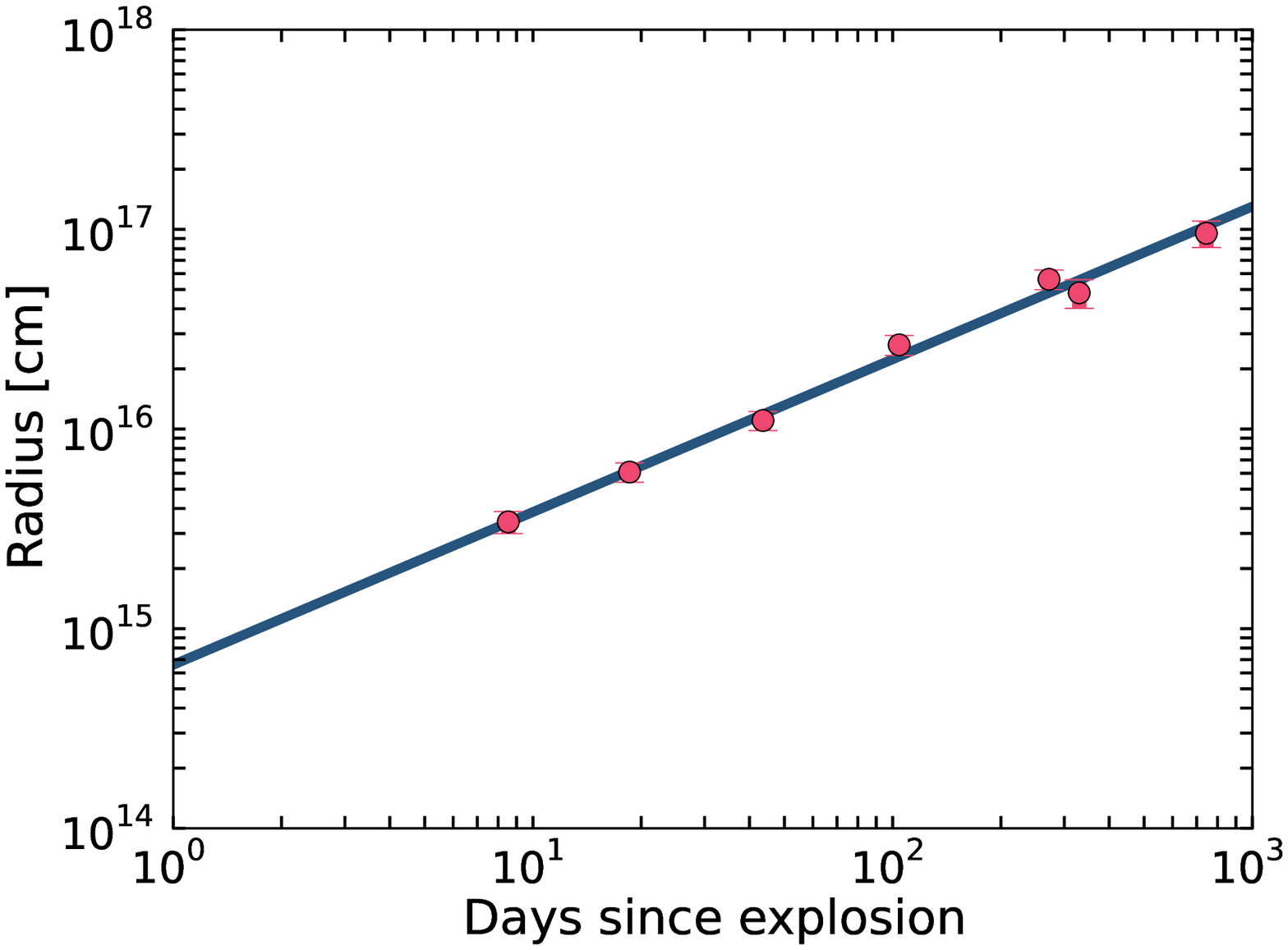}
    \includegraphics*[width=5.8cm]{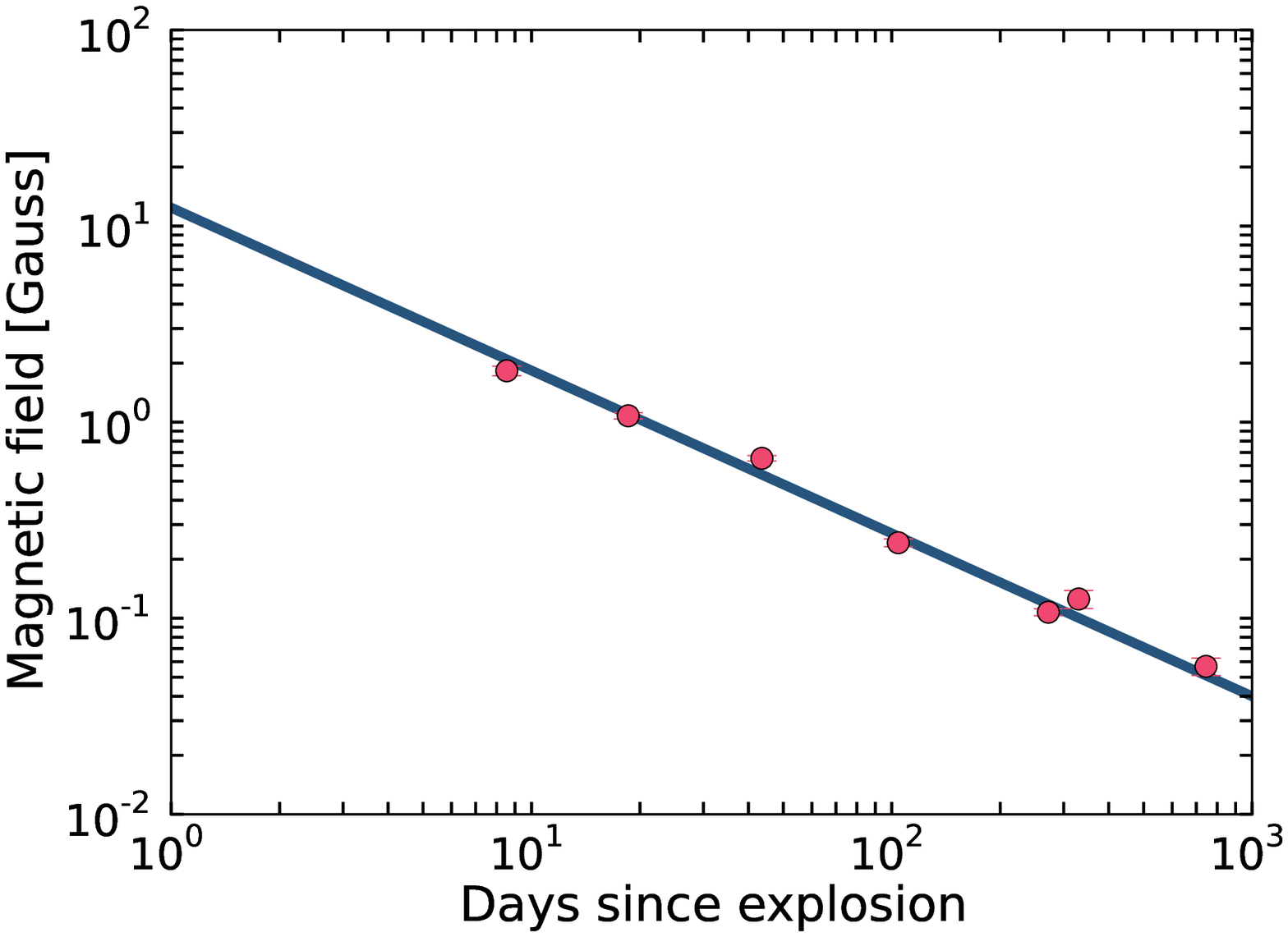}
    \includegraphics*[width=5.8cm]{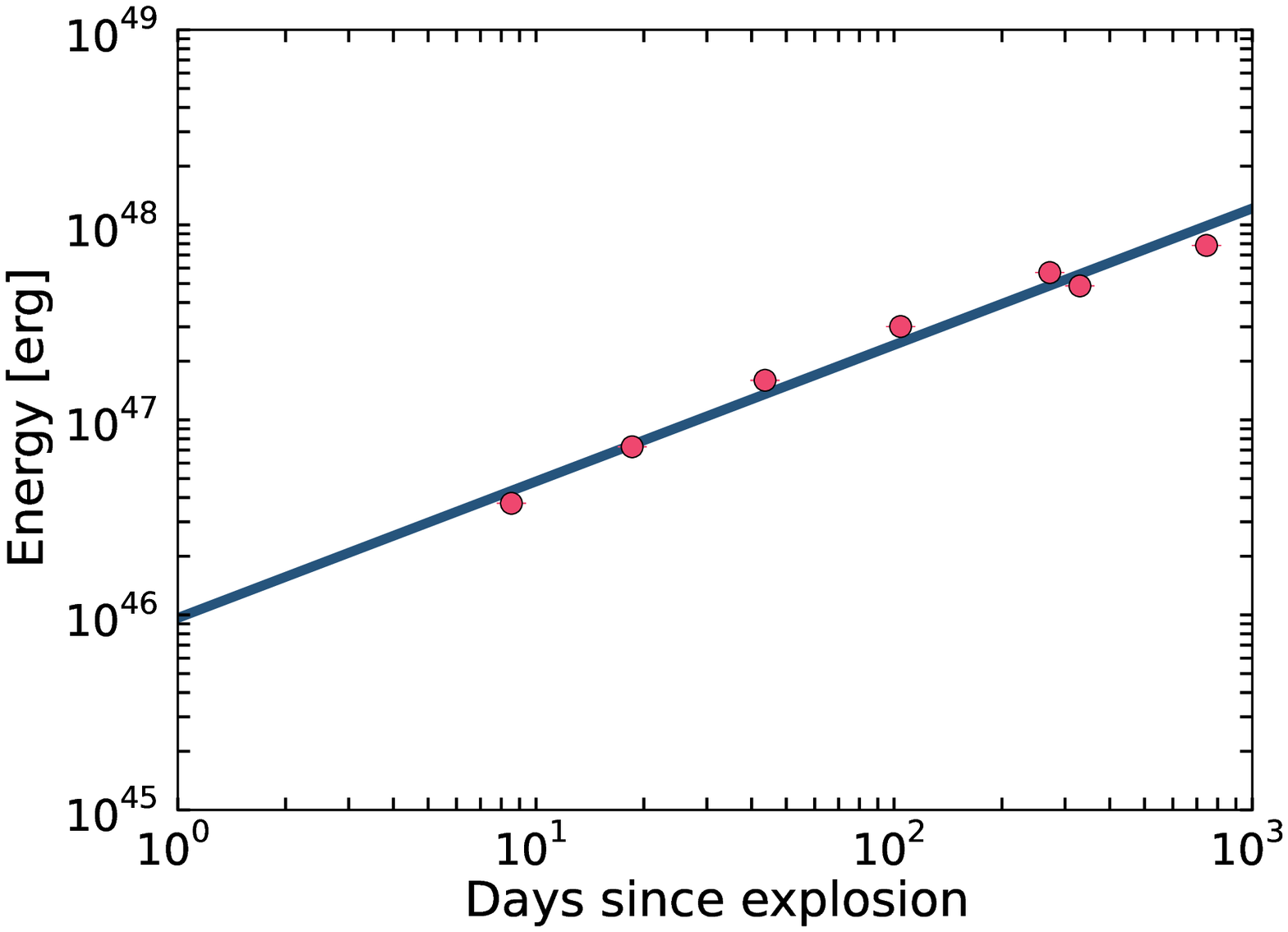}
\caption{The blast wave radius (left panel), magnetic field (middle panel) and minimum internal energy (right panel) estimates at seven epochs. The epochs are [8.54, 18.56, 43.60, 104.25, 272.00, 330.11, 745.01] days post explosion assuming the date of explosion to be 2016 May 25.9 (UT). The corresponding values are: R $\sim$ [0.34, 0.61, 1.10, 2.64, 5.62, 4.81, 9.56] $\times$ 10$^{16}$ cm, B $\sim$ [1.83, 1.08, 0.65, 0.24, 0.11, 0.13, 0.06] G and $E_{\rm min}$ $\sim$ [0.37, 0.73, 1.60, 3.01, 5.69, 4.86, 7.84] $\times$ $10^{47}$ erg. The solid blue curves are power-law fits with $R \propto$ t$^{0.77\pm0.03}$ (left panel), $B \propto$ t$^{-0.83\pm0.04}$ (middle panel) and $E_{\rm min}$ $\propto$ t$^{0.70\pm0.04}$ (right panel). The error bars in the figure (see table \ref{tab:blast wave para}) are smaller than the marker size.}
\label{fig:blast-wave-para}
\end{centering}
\end{figure*}

The spectra of ASASSN-16fp at multiple epochs are well represented by SSA spectrum as shown in Fig. \ref{fig:spectra-single-epoch-fit} with $p \sim 2.4$. The nearest $p$ value for which $c_{5}$ and $c_{6}$ are tabulated in \cite{pacholczyk1970} is for $p =2.5$. Hence we use $p=2.5$ and corresponding $c_{5}$ and $c_{6}$ values in eqns \ref{eqn:Rp-general} and \ref{eqn:Bp-general}. At later epochs, the spectrum peaks at lower frequencies due to the expansion of the blast wave. We find $\nu_{\rm p}$ $\sim$  20.49, 12.26, 07.66, 02.74 and 1.19 GHz and $F_{\rm p}$ $\sim$ 13.51, 15.37, 20.00, 14.39 and 12.20 mJy on $\sim$ day 8, 18, 43, 104 and 272 respectively. In addition to these five epoch spectra, we also have low frequency GMRT light curves (Fig \ref{fig:lc-gmrt}). The 1.39 and 0.61 GHz GMRT light curve peaks at $\sim$ 330.11 and 745.01 days post explosion respectively. The corresponding peak flux densities are $F_{\rm p}$ $\sim$ 12.18 and 9.12 mJy at 330.11 and 745.01 days post explosion respectively. We use $F_{\rm p}$ and $\nu_{\rm p}$ at these seven epochs to derive the blast wave radius and magnetic field strength. We also find the temporal evolution of these parameters by fitting a power-law to multi-epoch values of $R$ and $B$ independently. The SSA frequency cascades as $\nu_{\rm p}$ $\propto$ $t^{-0.85}$, consistent with radio SNe with dominant SSA \citep{soderberg2006,chevalier1998}. We also calculate the mean velocity of the radio emitting shell at each epoch as $R_{\rm p}/t$. The results are presented in Table \ref{tab:blast wave para} and Fig \ref{fig:blast-wave-para}.

The radius of the shock wave is $R_{1}$ = (0.34 $\pm$ 0.04) $\times$ 10$^{16}$ cm at 8.54 days post explosion and expands to $R_{2}$ = (9.56 $\pm$ 1.45) $\times$ 10$^{16}$ cm at 745.01 days post explosion. The temporal evolution of shock radius can be described as $R=3.4\times$10$^{15}$(t/8.54\,days)$^{0.77\pm0.03}$\, cm, indicative of a decelerating blast wave. The radial evolution of shock radius is slower compared to other type Ibc SNe like SN\,2003L \citep[$R \propto t^{0.96}$;][]{soderberg2005}, SN\,1983N \citep[$R \propto t^{0.86}$;][]{chevalier1998} and SN\,2007gr \citep[$R \propto t^{0.9}$;][]{soderberg2010a}. However the temporal index is within the expected range of values i.e 0.67 $\leq m \leq$ 1.0 \citep{chevalier1996,chevalier1998}. The post shock magnetic field evolves as $B=1.83(t/8.54\,\rm days$)$^{-0.83\pm0.04}$ with a radial dependence of $B \propto R^{-1.04\pm0.03}$\, G. The radial index of magnetic field is similar to that of other type Ic events like SN\,2003L \citep[$B \propto R^{-1.04}$;][]{soderberg2005} and SN\,2002ap \citep[$B \propto R^{-1};$][]{berger2002} The temporal index ($\alpha_{B}$) of magnetic field depends on the shock radius and CSM density as $\alpha_{B}$ = $[m(2-s$)/2]$-$1. Thus the derived $m$ and $\alpha_{B}$ imply the CSM density index to be $s=-1.56\pm0.06$. This implies that the CSM density of ASASSN-16fp is slightly flatter than the density field created by a steady stellar wind ($s = 2$). The derived values of $m$ and $s$ implies the ejecta density index of ASASSN-16fp to be $n = 7.83 \pm 0.4$.

The mean velocity of the shocked shell is $\sim$ 0.15c ($\sim$ 46443 km\,s$^{-1}$) on $\sim$ 8 days post-explosion indicative of a sub-relativistic velocity similar to that of normal type Ibc SNe \citep{chevalier2006}. 
%This is significantly lower than the ultra-relativistic or relativistic velocities seen in GRB associated SNe \citep{soderberg2010b}. This implies that the broad absorption features in the optical spectra as seen in ASASSN-16fp does not always indicate relativistic ejecta. 
If any other absorption process like FFA defines the peak of radio light curves, the actual mean velocities derived from optical measurements will be greater than the values derived from SSA model \citep{chevalier1998}. The optical line velocities of ASASSN-16fp derived from absorption features at roughly the same epoch is 35000 $\pm$ 10000 km\,s$^{-1}$ \citep{prentice2018}, which is less than the SSA derived value. Thus FFA is not likely to be the dominant absorption process in ASASSN-16fp.

\begin{table*}
    \centering
    \caption{Blast wave parameters of ASASSN-16fp.}
    \label{tab:blast wave para}
    \begin{tabular}{cccccc} % four columns, alignment for each
        \hline
    Days &  Blastwave radius & Magnetic field & Velocity & Mass-loss rate \\ 
 - &  ($\times$ $10^{16}$ cm) & ($\times$ 10$^{-1}$ Gauss) &  ($\times$ c) & ($\times$10$^{-5}$ $M_{\odot}\rm yr^{-1}$) \\
        \hline
8.54   & 0.34 $\pm$ 0.04  & 18.27 $\pm$ 1.02 & 0.15 $\pm$ 0.02 & 0.44 $\pm$ 0.16\\
18.56  & 0.61 $\pm$ 0.07  & 10.78 $\pm$ 0.40 & 0.13 $\pm$ 0.02 & 0.71 $\pm$ 0.23\\
43.60  & 1.10 $\pm$ 0.12  & 6.54 $\pm$ 0.21 & 0.10 $\pm$ 0.01 & 1.45 $\pm$ 0.46 \\
104.25  & 2.64 $\pm$ 0.30  & 2.43 $\pm$ 0.11 & 0.10 $\pm$ 0.01 & 1.14 $\pm$ 0.38 \\
272.00  & 5.62 $\pm$ 0.63  & 1.07 $\pm$ 0.04 & 0.08 $\pm$ 0.01 & 1.52 $\pm$ 0.50\\
330.11  & 4.81 $\pm$ 0.79  & 1.25 $\pm$ 0.01 & 0.06 $\pm$ 0.01 & 3.06 $\pm$ 1.56 \\
745.01  & 09.56 $\pm$ 1.45 & 0.57 $\pm$ 0.01 & 0.05 $\pm$ 0.01 & 3.20 $\pm$ 1.52 \\
        \hline
    \end{tabular}\\
\parbox{100mm}{
The blast wave radius (R), magnetic field strength (B), mean shell velocity and mass-loss rates at seven epochs, day 8.54, 18.56, 43.60, 104.25, 272.00, 330.11 and 745.01 post explosion. We assume the date of explosion as 2016 May 25.9 (UT).
}
\end{table*}

\subsection{Internal energy of the radio emitting material}
The minimum total internal energy of the ejecta to power the observed radio emission can be found from magnetic energy density \citep{soderberg2010a}.
\begin{equation}
E_{\rm min} = \frac{1}{\epsilon_{\rm B}} \frac{4}{3} \pi R^{3}\, f \frac{B^{2}}{8 \pi}
\end{equation}
Assuming equipartition of energy between relativistic electrons and magnetic fields, we take $\epsilon_{\rm e}$ = $\epsilon_{\rm B}$ = 0.33. This places a lower limit to the total internal energy of radio emitting medium \citep{soderberg2010a}. 

\begin{equation}
E_{\rm min} \approx 2.78 \times 10^{44} \left( \frac{B}{1\, \rm G} \right)^{2} \left( \frac{R}{10^{15} \rm cm} \right) ^{3}
\end{equation}
For the derived parameters of $B$ and $R$, we calculate $E_{\rm min}$ = [0.37$\pm$0.15, 0.73$\pm$0.25, 1.60$\pm$0.53, 3.01$\pm$1.07, 5.69$\pm$1.97, 4.86$\pm$2.60, 7.84$\pm$3.91] $\times$ $10^{47}$ erg on day 8.54, 18.56, 43.60, 104.25, 272.00, 330.11 and 745.01 post explosion respectively (see Table \ref{tab:blast wave para}), comparable to other type Ibc SNe \citep{margutti2014}. The temporal evolution of the energy can be described as $E$ $\sim$ 0.37 $\times$ $10^{47}$ (t/8.54\,days)$^{0.65}$ erg (see Fig \ref{fig:blast-wave-para}), obtained from the temporal indices of $R$ and $B$. Any additional absorption process like FFA or non-equipartition values of $\epsilon_{\rm e}$ and $\epsilon_{\rm B}$ will further increase the energy of the radio emitting material \citep{chevalier2006,fransson1998}.

\begin{table*}
    \centering
    \caption{Comparison of ASASSN-16fp parameters with other radio bright SNe\,Ic-BL.}
    \label{comparison-typeIc-BL}
    \begin{tabular}{ccccccccccccc} % four columns, alignment for each
        \hline
SN & Distance & $t_{\rm p}^{a}$ & $\nu_{\rm p}^{a}$ & $F_{\rm p}^{a}$ & $L_{\rm 5\, \rm GHz}$ & B & E & $\dot{M}$ & m$^{b}$ & $R_{\rm p}/t_{\rm p}$ & References \\ 
- & (Mpc) & (days) & (GHz) & (mJy) & (erg s$^{-1}$ Hz$^{-1}$) & (G) & (erg\,s$^{-1}$) & ($10^{-6} M_{\odot}\rm yr^{-1}$) & - & ($\times$c) & - \\
        \hline
SN\,2002ap & 7.3 & 7 & 1.4 & 0.3 & 3.0$\times$10$^{25}$ & 0.26 & 1.5$\times$10$^{45}$ & 0.5 & 0.90 & 0.35 & 1,2,3 \\
SN\,2009bb & 40.0 & 20 & 6.0 & 19.0 & 1.9$\times$10$^{27}$ & 0.45 & 1.3$\times$10$^{49}$ & 2.0 & 0.94 & 0.85  & 4,5,6 \\
SN\,2012ap & 40.0 & 18 & 8.9 & 5.7 & 8.0$\times$10$^{27}$ & 1.08 & 1.6$\times$10$^{49}$ & 3.5 & 0.74 & 0.30 & 7,8,9,10  \\
PTF11qcj  & 124.0 & 10 & 1.0 & 0.7 & 1.0$\times$10$^{29}$ & 6.70 & 9.3$\times$10$^{48}$ & 120.0 & 0.80 & 0.42 & 11,12,13\\                                                                                              
% PTF11cmh  & 491 & 10 &  &     & 1.0$\times$10$^{29}$ & 5.0 & 6$\times$10$^{48}$ & 100 & 0.86 & 0.30 & 10 \\
 PTF14dby & 337.0 & 10 & 24.0 & 0.2   & 2.6$\times$10$^{28}$ & 1.60 & 8.0$\times$10$^{47}$ & 5.0 & 0.78 & 0.38 & 14,15,16 \\
 {\bf ASASSN-16fp$^{c}$} & {\bf 17.2} & {\bf 18} & {\bf 12.3} & {\bf 15.4} & {\bf 5.2$\times$10$^{27}$} & {\bf 1.08} & {\bf  0.7$\times$10$^{47}$} & {\bf 7.1} & {\bf 0.77} & {\bf 0.13} & 17,18,19   \\
 SN\,1998bw & 38.0 & 16 & 6.0 & 40.0 & 8.0$\times$10$^{28}$ &  0.40 & 1.0$\times$10$^{49}$ & 0.25 & 0.77 & 1.30 & 20,21,22                   \\
        \hline
\end{tabular}\\
\parbox{163mm}{
References: (1) \cite{berger2002}, (2) \cite{smartt2002}, (3) \cite{galyam2002}, (4) \cite{soderberg2010b}, (5) \cite{pignata2009}, (6) \cite{pignata2011}, (7) \cite{chakraborti2015}, (8) \cite{margutti2014}, (9) \cite{springob2007}, (10) \cite{milisavljevic2012}, (11) \cite{corsi2014}, (12) \cite{bloom2012}, (13) \cite{palliyaguru2019}, (14) \cite{yaron2012}, (15) \cite{laher2014}, (16) \cite{corsi2016}, (17) \cite{yamanaka2017}, (18) \cite{holoien2016}, (19) \cite{elias2016}, (20) \cite{kulkarni1998}, (21) \cite{li1999}, (22) \cite{galama1998}.\\
Note - Radio luminosity ($L_{\rm 5\,GHz}$) is peak spectral luminosity at $\sim$ 5\,GHz. The time to reach $L_{\rm 5\,GHz}$ for SN\,2002ap, SN\,2009bb, SN\,2012ap, PTF11qcj, PTF14dby, ASASSN-16fp and SN\,1998bw are 3, 52, 38, 100, 47, 104 and 12 days post explosion respectively.\\
Note - The parameters $B$, $E$, $\dot{M}$ and $R_{\rm p}/t_{\rm p}$ are derived at the time of SSA peak assuming equipartition of energy between relativistic electrons and magnetic field.\\
$a$ \, $\nu_{\rm p}$, $F_{\rm p}$ and $t_{\rm p}$ denotes the peak frequency, peak flux density and time to peak respectively of the SSA modelled spectra \citep{chevalier1998}.\\
$b$ \, $m$ denotes the shock deceleration parameter. \\
$c$ \, The parameters of ASASSN-16fp are from this work.
}
\end{table*}

\subsection{Mass loss rate of the progenitor star}

%We use the blast wave parameters derived on multiple epochs to estimate the mass-loss rate of the progenitor star, assuming energy equipartition between magnetic field and relativistic electrons. 
The mass-loss rate of the progenitor star can be derived from the post shock magnetic field energy density \citep{chevalier1998,soderberg2006}.

\begin{equation}
U_{\rm B} = \frac{B^2}{8\pi} \approx \frac{\epsilon_{B}}{4\pi}\frac{\dot{M}}{v_{\rm w}}R^{-2}v^2
\end{equation} 
where $v_{\rm w}$ is the velocity of the stellar wind of the progenitor star. Assuming $\epsilon_{B}$ = 0.33 and a wind velocity of $v_{\rm w}$ $\sim$ 1000 km\,s$^{-1}$, typical of a WR star \citep{cappa2004}, we derive the mass-loss rate to be, $\dot{M}$ $=$ $(0.44\pm0.16)$  $\times$ 10$^{-5}$ $M_{\odot}\rm yr^{-1}$ on day 8.54 post explosion. The mass-loss rate is $\dot{M}$ $=$ $(3.20\pm1.52)$  $\times$ 10$^{-5}$ $M_{\odot}\rm yr^{-1}$ on day 745 post explosion (see Table \ref{tab:blast wave para}). 
 The mass-loss rates at multiple epochs suggest that the progenitor of ASASSN-16fp has gone through variable mass-loss rates in the years prior explosion. The mass-loss rate at $\sim$ 37 years prior explosion is 7.3 times greater than the mass-loss rate at $\sim$ 1 year prior explosion for a stellar wind velocity of 1000 km\,s$^{-1}$. The derived mass-loss rates are consistent with the mass-loss rate seen in Galactic WR stars ($1-5$) $\times$ $10^{-5}$ $M_{\odot} \rm yr^{-1}$\citep{abbott1986,leitherer1995,leitherer1997,chapman1999,cappa2004}. 
 
The equipartition assumption puts a lower limit on the mass-loss rate. In a realistic scenario, the SN post shock energy is distributed among electrons, protons/ions and magnetic fields and the values of $\epsilon_{\rm B}$ and $\epsilon_{\rm e}$ are likely less than 0.33 \citep{chevalier2006}. For more realistic values, $\epsilon_{\rm B}=0.01$ and $\epsilon_{\rm e}=0.1$ \citep{terreran2019}, we derive the mass-loss rates to be $\dot{M}$ $=$ ($0.5-3.8$) $\times$ $10^{-4}$ $M_{\odot} \rm yr^{-1}$ at various epochs spanning $8-745$ days post-explosion. This is consistent with the mass-loss rate estimate of ASASSN-16fp $\dot{M}$ $=$ ($1-2$) $\times$ $10^{-4}$ $M_{\odot} \rm yr^{-1}$ from X-ray observations \citep{terreran2019}.

\subsection{Density structure of CSM}
\label{sec:density-variations}
Radio light curves trace the density structure of the CSM. The density of the CSM need not be uniform due to variable mass-loss rate, variable wind velocity of the progenitor star, ejected stellar envelopes of progenitors, etc. The non-uniform density structure of the CSM can cause jumps in the radio light curve \citep{soderberg2006}. In the radio light curve of ASASSN-16fp, we see flux density enhancements on day 43.6 post-explosion at frequencies above $\sim$ 10 GHz where the emission is optically thin. The effect is also seen in the spectra (see Fig. 5) where all the optically thin flux density measurements are above the model prediction. We interpret this as a signature of non-uniform CSM density at a radius of $\sim$ 1.10 $\times$ 10$^{16}$ cm. The flux density scales as $R^{3} N_{0} B^{(p+1)/2}$, for SSA dominated radio emission \citep{chevalier1998}. Assuming constant $\epsilon_{B}$ through out the evolution, $B^{2}$ $\propto$ $n_{e}v^{2}$ and the radio flux density $F_{\nu} \propto n_{e}^{(p+5)/4}$. For $p=2.4$, $F_{\nu} \propto n_{e}^{1.9}$. The flux density enhancement seen on day 43 from ASASSN-16fp is $\sim$ 1.5 times that of the standard model prediction. Thus the density enhancement in CSM is only $\sim$ 1.2. This could be due to small scale clumping within the stellar wind \citep{moffat2008,smith2014}. The CSM of WR stars are known to be significantly disturbed and there are X-ray observations that show evidence for dense clumps \citep{hillier2003}. Assuming a stellar wind velocity of 1000 km\,s$^{-1}$, typical of WR stars \citep{cappa2004}, the density enhancement could be due to a mass-loss event happened $\sim$ 3.5 years prior explosion. These sorts of small scale flux density enhancements are seen in $\sim$ 50\% of SNe Ib/c \citep{soderberg2006}.

\section{A comparison with other SNe Ic-BL}
In this section, we compare the properties of ASASSN-16fp with other radio bright SNe\,Ic-BL with out GRB association. The properties of a few SNe\,Ic-BL from radio modeling are compiled in Table \ref{comparison-typeIc-BL}. We also include the first prototypical GRB associated SN\,1998bw/GRB980425 in the table for comparison. The 5 GHz light curve of ASASSN-16fp peaks at $\sim$ 104 days post-explosion with spectral luminosity 5.2$\times$10$^{27}$ placing it as one of the luminous radio SNe with luminosities similar to SN\,2009bb \citep{soderberg2010b} and SN\,2012ap \citep{chakraborti2015}. The peak radio spectral luminosity of ASASSN-16fp is within the broad distribution of radio luminosities of SNe\,Ibc \citep{soderberg2006-68Ibc} attributed to the range of CSM densities observed in Galactic WR stars \citep{chevalier2006}. Besides, the variation in the parameters $\epsilon_{\rm e}$ and $\epsilon_{\rm B}$ may also contribute to the large range of radio luminosity \citep{chevalier2006}. The light curve of ASASSN-16fp evolves slightly slower compared to the typical rise time of 5 GHz light curve (10-30 days) of SNe\,Ib/c \citep{weiler1998}. The 5 GHz spectral luminosity of ASASSN-16fp is 20 times smaller than the spectral luminosity of PTF11qcj \citep{corsi2014,corsi2016} that peaks at a similar time (100 days post-explosion). The mass-loss rate of PTF11qcj is $\sim$ 17 times more than that of ASASSN-16fp owing to the brighter radio emission possibly due to denser CSM. The mean shock velocity of ASASSN-16fp on $\sim$ 18 days post-explosion is 0.13c consistent with the mean shock velocities of normal SNe\,Ib/c \citep[0.1-0.15c;][]{soderberg2006-68Ibc}. \cite{chevalier1998} compiled mean shock velocities of a sample of SNe\,Ib/c and the mean shock velocity of ASASSN-16fp closely resemble the shock velocity of SN\,1993N at similar epoch. A comparison with the mean shock velocities of SNe\,Ic-BL suggests that ASASSN-16fp has the slowest shock with velocity at least a factor of 2 less than the rest of the SNe\,Ic-BL in the Table \ref{comparison-typeIc-BL}. The shock velocity of ASASSN-16fp is $\sim$ 10 times slower compared to GRB associated SN\,1998bw \citep{kulkarni1998}. Thus ASASSN-16fp further emphasizes that broad lines in the optical spectra cannot be considered as the proxy for relativistic ejecta as also seen in SN\,2002ap \citep{berger2003}. The difference in shock velocity could be either due to the difference in CSM density or due to the SN property itself like initial explosion energy or ejecta mass. The different SNe\,Ic-BL are characterized by roughly similar shock deceleration parameter ($m$). The shock deceleration parameter of ASASSN-16fp is $m \sim 0.8$, indicative of WR like radiative progenitor star similar to PTF11qcj \citep{corsi2014} and PTF11dby \citep{corsi2016,horesh2013}. The magnetic field of ASASSN-16fp at $\sim$ 18 days post-explosion (1.1 G) is higher than the magnetic fields seen in normal SNe\,Ib/c \citep[0.2-0.6 G;][]{chevalier1998} and is similar to the magnetic field of SN\,2012ap \citep{chakraborti2015}. The mass-loss rates of the progenitors of SNe\,Ic-BL are in the range of (0.3-5)$\times$10$^{-6}$ $\dot{M}$\,yr$^{-1}$ except for PTF11qcj \citep{corsi2014}. The relatively lower mass-loss rates are indicative of compact progenitors of these SNe driving faster stellar winds. The internal energy of the radio-emitting material of ASASSN-16fp is 46 times greater than SN\,2002ap \citep{berger2003} and $\sim$ 100 times smaller than SN\,1998bw \citep{kulkarni1998}, placing it in a phase space between SN\,2002ap and SN\,1998bw in terms of energetics. The radio light curve of ASASSN-16fp shows remarkable similarity with the radio light curves of PTF11qcj \citep{corsi2014} showing achromatic short time scale variability \citep[see Fig 11 and 12 of][]{corsi2014}. PTF11qcj exhibits a factor of 2 enhancement in flux density at radius $>$1.7$\times$10$^{17}$ cm owing to small-scale density fluctuations in the CSM. ASASSN-16fp also shows small scale flux density enhancement at radius $R$=1.1$\times$10$^{16}$ cm (see \S \ref{sec:density-variations}).

\section{Summary}
\label{sec:summary}
We present extensive radio observations of a Type Ic supernova, ASASSN-16fp spanning a frequency range of $0.33-25$ GHz and a temporal range of $\sim$ $8-1136$ days post-explosion. We model the radio data with the standard model and our main results are the following
\begin{enumerate}
\item The observations are best represented by a model in which the dominant absorption process is SSA during the time-scale probed by the radio data. 
 \item Assuming equipartition of energy between relativistic particles and magnetic fields ($\epsilon_{\rm e} = 0.33$ and $\epsilon_{\rm B}=0.33$), we estimate the shock radius and velocity to be $R \sim 0.34 \times 10^{16}$ cm and $v \sim 0.15$c respectively at $t_{0} \sim$ 8 days post-explosion. The shock velocity is sub-relativistic as seen in normal type Ic SNe implying that the broad absorption lines in the optical spectra do not indicate relativistic ejecta.
 
\item The evolution of the shock radius and magnetic field can be represented as $R \propto$ t$^{0.77\pm0.03}$ and $B \propto$ t$^{-0.83\pm0.04}$ respectively, implying a CSM density profile $\rho_{\rm csm} \propto r^{-1.6}$ and an outer ejecta density profile $\rho_{\rm ej} \propto r^{-8}$.
  
\item  We infer the temporal evolution of the total internal energy of the radio-emitting material to be $E$ $\sim$ 0.37 $\times$ $10^{47}$ (t/8.54\,days)$^{0.65}$ erg, consistent with the normal type Ibc SN population. 

 \item We determine the mass-loss rate of the progenitor star to be $\dot{M}$ $\sim$ $(0.4-3.2) \times10^{-5}$ $M_{\odot}\rm yr^{-1}$ from equipartition values, consistent with the mass-loss rate of Galactic \textbf{WR} stars. 
 %Assuming $\epsilon_{\rm B}=0.01$ and $\epsilon_{\rm e}=0.1$, the mass-loss rate is $\dot{M}$ $=$ ($0.5-3.8$) $\times$ $10^{-4}$ $M_{\odot} \rm yr^{-1}$ roughly consistent with the mass-loss rate estimates from X-ray observations \citep{terreran2019}.
  
 \item The radio light curves and spectra show the signature of density enhancement in the CSM at a radius of $\sim$ $1.1 \times 10^{16}$ cm from the explosion center possibly due to a small scale clumping in the stellar wind $\sim$ 3.5 years prior explosion.
 
 \item A comparison of ASASSN-16fp parameters with other SNe\,Ic-BL suggests that ASASSN-16fp is fairly radio luminous similar to SN\,2012ap and SN\,2009bb with slower shock typical of normal SNe\,Ibc.

\end{enumerate}
\section*{Acknowledgements}
P.C. acknowledges support from the Department of Science and Technology via SwaranaJayanti Fellowship award (file no.DST/SJF/PSA-01/2014-15). We thank the staff of the GMRT that made these observations possible. GMRT is run by the National Centre for Radio Astrophysics of the Tata Institute of Fundamental Research.
The National Radio Astronomy Observatory is a facility of the National Science Foundation operated under cooperative agreement by Associated Universities, Inc.

%%%%%%%%%%%%%%%%%%%%%%%%%%%%%%%%%%%%%%%%%%%%%%%%%%



%\bibliographystyle{mnras}

%%%%%%%%%%%%%%%%%%%%%%%%%%%%%%%%%%%%%%%%%%%%%%%%%%

%%%%%%%%%%%%%%%%% APPENDICES %%%%%%%%%%%%%%%%%%%%%

%\appendix



%%%%%%%%%%%%%%%%%%%%%%%%%%%%%%%%%%%%%%%%%%%%%%%%%%

% Don't change these lines
\bsp    % typesetting comment
\label{lastpage}
\end{document}